# A discrete ordinates method of radiative transfer in one-dimensional spherical geometry

Charles H Aboughantous[*]

*Louisiana State University, Department of Physics and Astronomy, Baton Rouge, Louisiana 70803, USA*

**Abstract.** A new set of discrete ordinates is proposed for one-dimensional radiative transfer in spheres with central symmetry. The set is structured with un-normalized circular functions. This resulted in a conservative and closed set of discrete ordinates equations that need not starter intensity: $N$ equations with $N$ unknowns. The angular derivative of the transfer equation is represented by a set of angular parameters that are calculated directly and without recursion from the abscissas and weights of the quadrature and without the constraint of asymptotic conditions. An analytic solution is obtained in infinite homogeneous and heterogeneous cold media for monochromatic radiation. Graphical and tabulated data validate the proposed set of discrete ordinates at all radial positions ranging from $10^{-7}$ to $10^{+7}$ within round-off errors.

## 1. Introduction

Extensive efforts were devoted to the discrete ordinates method in radiation transport, which was successfully coupled with analytical solutions on the spatial domain in slab geometry [1] [2]. Fewer efforts were devoted to produce analytical solutions in discrete ordinates formalism in spherical geometry. Chandrasekhar [1] proposed an analytic approach based on the $P_n$ approximation in discrete directions. In order to obtain the solution for the transfer equation, he expanded the specific intensity in $P_4$ and applied a sequence of transformations to the moments of the specific intensity in specific directions defined by the abscissas of Gauss-Legendre (GL) quadrature. These transformations yielded a Lommel equation for an accessory function. The solution was then converted to the total intensity, the first and the second moments. This approach is tedious and unyielding to expansions for $n > 4$.

Further developments in the discrete ordinates formalism in spherical geometry were made in the realm of numerical methods. These efforts led to a set of discrete ordinates structured from *angular parameters* that approximate the angular derivative. The angular parameters are computed recursively from the abscissas and weights of GL quadrature constrained to a constant asymptotic specific intensity [3]. The resulting set of equations is not closed: there are $N$ linear equations with $N+1$ unknowns. The additional unknown, labeled the *starter intensity*, is borrowed from the solution of the transfer equation in slab geometry with a *starter direction cosine* $\mu = -1$. This discrete ordinates model produces a *dip* at the center of the sphere. It was believed that the dip is not physical and recipes were proposed to cure it.

One approach to cure the dip evolves around applying Radeau quadrature in the first angular cell and GL quadrature in the interior angular cells using the same structure of angular parameters. The dip was almost cured with micrometric spatial cells [4]. An alternative approach was proposed where the starter direction cosine has to be found experimentally. It was found that this starter direction cosine depends on the radius of the sphere, on the spatial grid and its magnitude can be larger than 1 in some spheres [5].

---
[*] email address: abough@rouge.phys.lsu.edu

A recent approach complemented the condition of constant specific intensity at the center by requiring an isotropic intensity at $r = 0$. For the most part the GL quadrature is replaced by an angular difference $\Delta\mu$ along with a new set of interpolating functions representing the specific intensity in the angular cells. This approach appears to cure the dip at the center of the sphere but a starter intensity is still needed [6]. The shortcoming of this approach is that the finite and isotropic intensity at the center is not a physical boundary condition and the starter intensity of slab geometry is not valid particularly in the interior of spherical media. It was shown that this boundary condition defines a boundary value problem different from the problem of energy flow in spheres [7] and the starter flux should be consistent with the inverse square law [8].

We revisited the structure of the set of discrete ordinates in one-dimensional spherical geometry and proposed a new approach presented in this paper. The new set is based on GL quadrature. The angular derivative is discretized with a finite set of angular parameters derived directly from the abscissas and weights of the quadrature and are not constrained to an asymptotic isotropic intensity. Consequently, the set of discrete ordinates equations is closed and there is no need for starter intensity. This is a major departure from the original discrete ordinates method.

In order to validate the structure of the proposed set of discrete ordinates, we carried out the general analytic solution continuous in $r$ in a homogeneous infinite medium and in end-points form suitable for heterogeneous infinite cold media. The latter solution is in fact a numerical algorithm free of spatial truncation errors. Considering that the approximation introduced by discrete ordinates is indeed an approximation of the divergence term of the transfer equation, the validity of the proposed discrete ordinates set is demonstrated in vacuum. The import of the approximation to the solution in cold opaque media is also examined quantitatively.

## 2. Definition of the problem

Consider a sphere of radius $R$ embedding a concentric *pellet* of small radius $\varepsilon < R$. The pellet and the sphere may or may not be thermodynamically identical. Whichever is the case we treat them as two distinct media. Considering that transfer equation object of this paper is the linearized form of the transport equation, this equation is not valid in the proximity and at the center. Therefore, intensities are not calculated in the pellet region but specular boundary condition is applied at its surface. The sphere is assumed homogeneous and isothermal, characterized by its black body constant $B$ and a plankian opacity $\kappa$; it will be stratified to simulate a heterogeneous medium in a later section.

The sphere is assumed at LTE, characterized by emission density $\kappa B$ and is immersed in a

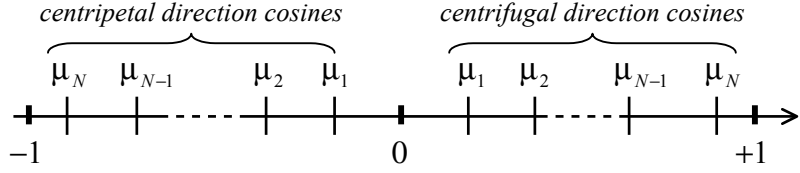

**Figure 3.1.** Layout of the angular grid for the centripetal and the centrifugal direction cosines. The boundary values ±1 are references to the actual boundary limits for the direction cosines. None of the direction cosines should be construed as a negative quantity.

uniform and isotropic radiation field of specific intensity $\psi_R^-$. Hence, the transfer equation without scattering may be written in conservation form as:

$$\frac{\mu}{r^2}\partial_r(r^2\psi_r) + \frac{1}{r}\partial_\mu(\eta^2\psi_r) + \kappa\psi_r = \kappa B \qquad r \in [\varepsilon, R] \qquad (2.1)$$

where $\partial_r$ is the tensor notation for the derivative with respect to $r$, $\psi$ is the specific intensity, $\mu$ the direction cosine on $[-1, +1]$ and $\eta^2 = (1 - \mu^2)$. The subscript in $\psi_r$ is for the spatial variable; the angular variable is omitted, it is implied by the mere use of the letter $\psi$. We used the symbol $\psi$ for the specific intensity to avoid confusion with the traditional symbol $I$ reserved for the identity matrix. The search for the solution of (2.1) is greatly simplified using Schuster's technique [9] of the two streams representation on $\mu \in [0, 1]$ and $r \in [\varepsilon, R]$:

*Centripetal equation*:
$$\frac{\mu}{r^2}\partial_r(r^2\psi_r^-) + \frac{1}{r}\partial_\mu(\eta^2\psi_r^-) - \kappa\psi_r^- = -\kappa B \qquad (2.2)$$

*Centrifugal equation*:
$$\frac{\mu}{r^2}\partial_r(r^2\psi_r^+) + \frac{1}{r}\partial_\mu(\eta^2\psi_r^+) + \kappa\psi_r^+ = \kappa B \qquad (2.3)$$

The mathematical boundary conditions are $\psi_\varepsilon^-$ and $\psi_\varepsilon^+$ at the surface of the pellet. They will be used to initiate the solutions and subsequently will be resolved in terms of the natural boundary condition $\psi_R^-$ by appropriate inversion of the solution of the centripetal equation.

## 3. The set of discrete ordinates

We define our set of discrete ordinates using the traditional set of weights and abscissas of GL quadrature:

$$S_{2N} \equiv \{w_n, \mu_n : w_n \in (0, 1), \mu_n \in (0, 1), n = 1, 2, \ldots, N\} \qquad (3.1)$$

We label the ordinates so that $\mu_1 < \ldots < \mu_N < 1.0$ and we order them in the sequence shown in figure 3.1. By this arrangement, larger $n$ implies larger direction cosine and $\mu_n > \mu_{n-1}$ is always true in either stream. All direction cosines are positive as specified by the set $S_{2N}$ and the weights are normalized so that: $\Sigma_1^N w_n = 1$. A subtle variation of this set was previously adopted by Feautrier [10] but in a different context.

We define a *discrete ordinates derivative* the operator $\delta/\delta\mu_n$ where $\delta\mu_n$ is a *differential entity* in $S_{2N}$ about the direction cosine $\mu_n$, conceptually different from the traditional finite difference $\Delta\mu = \mu_n - \mu_{n-1}$. We find the worth of $\delta\mu_n$ from the very definition of Gauss quadrature: if $f(\mu)$ is a polynomial of degree no greater than $4N-1$, then at some point $\mu_n \in [a, b] \subset (-1, +1)$ we must have exactly:

$$w_n f(\mu_n) \equiv w_n f_n = \int_a^b f(\mu) \, d\mu \qquad (3.2)$$

It follows that if we wish to integrate the discrete ordinates derivative of our function $f(\mu)$ on the same domain $[a, b]$, we must have:

$$w_n \left.\frac{df}{d\mu}\right|_{\mu_n} = w_n \frac{\delta f_n}{\delta \mu_n} = \int_a^b df = f_b - f_a \qquad (3.3)$$

It is apparent from (3.3) that if we assign the value $f_b - f_a$ to $\delta f$, then it is imperative to assign the weight $w_n$ to $\delta \mu_n$. Therefore, for gaussian discrete ordinates that satisfy identity (3.3) we must have:

$$\delta \mu_n = w_n \quad \Leftrightarrow \quad \delta f_n = f_n - f_{n-1} \qquad (3.4)$$

It follows that the angular derivative of $f$ transforms into a discrete ordinates derivative by the representation:

$$\left.\frac{\partial f}{\partial \mu}\right|_{\mu_n} \rightarrow \frac{f_n - f_{n-1}}{w_n} \qquad (3.5)$$

If we choose the function $f$ of identity (3.2) to be the angular derivative of (2.1), then (3.3) yields:

$$w_n \frac{\delta[\eta^2 \psi]_n}{\delta \mu_n} = \delta[\eta^2 \psi]_n = \eta_n^2 \psi_n - \eta_{n-1}^2 \psi_{n-1} \qquad (3.6)$$

which is the proper quadrature of the angular derivative on the specified domain.

There is still one bump in the way of the quadrature of the angular derivative: the closure under quadrature: the quadrature of the angular derivative of (2.1) must match the integral of the same:

$$\sum_{n=1}^{2N} \left( w_n \frac{\delta[\eta^2 \psi]_n}{\delta \mu_n} \right) = \int_{-1}^{1} \partial_\mu [\eta^2 \psi] \, d\mu \qquad (3.7)$$

It is readily apparent that the integral of the right side of (3.7) vanishes and, therefore, the sum of the left side must vanish as well. Expand the summation of (3.7) following the indices pattern shown in figure 3.1 and make use of identity (3.6), all the terms cancel out except those affected by the index $N$. Equating this result with the vanishing integral of (3.7) yields:

$$\eta_N^2 \left( \psi_N^+ + \psi_N^- \right) = 0 \qquad (3.8)$$

We recognize that in the asymptotic limits as $N \rightarrow \infty$ the identity (3.8) vanishes by reason of asymptotic direction cosine $\mu_\infty = 1 \Rightarrow \eta_\infty = 0$. It follows that for any finite $N$ the nullity of identity (3.8) must be postulated because $\eta_N$ is never equal to zero. To this end, the value $\mu_N = 1$ must be excluded simply because this value of direction cosine is not from the set $S_{2N}$ defined by (3.1). A direct approach to resolve this paradox is to work with un-normalized circular functions. We refer to this as *discrete ordinates trigonometry* (DOT): the circular functions on the unit circle map onto circular functions on the DOT circle. To accomplish this we choose the radius of our trigonometric circle equal to $\mu_N$ and then discretize the circular functions on the new circle. Therefore, the following transformation of the sine function applies:

$$\left.\begin{array}{l}\text{Unit circle}\\ \eta^2 = 1-\mu^2\end{array}\right\} \rightarrow \left\{\begin{array}{l}\text{DOT Circle}\\ \eta_n^2 = \mu_N^2 - \mu_n^2\end{array}\right. \Rightarrow \eta_N = 0 \qquad (3.9)$$

The set of discrete ordinates defined by (3.1) along with identities (3.2) and transformation (3.9) is complete. It is apparent from (3.9) that $\mu_N$ is the *nominal* normal direction cosine for the radial direction of radiation in $S_{2N}$.

Another thing we took care of when we expanded the summation (3.7): the discrete ordinates derivative about $\mu_1$. In accordance with (3.2) we would have to write:

$$\delta\eta^2\psi\big|_1 = \eta_1^2\psi_1 - \eta_0^2\psi_0 \qquad (3.10)$$

Considering that $\eta_n$ maps onto the set $S_{2N}$ by the circular relation (3.9), this mapping does not exist for $n = 0$. This is because the index $n = 0$ does not belong to the set $S_{2N}$ defined by (3.1). Therefore, if $\delta\eta\,\psi|_1$ is to be defined on $S_{2N}$, the term in $\eta_0$ of identity (3.10) must be dropped out notwithstanding any other considerations. With that, we will be left with only one significant term: $\delta(\eta\psi)\big|_1 = \eta_1^2\psi_1$. By doing so the integrity of the definition of the discrete ordinates derivative is preserved. The quantity $\delta\mu_1 = w_1$ is still completely specified about $\mu_1$ on $S_{2N}$.

Finally, the specific intensities are summed by quadrature from the set $S_{2N}$:

$$f_\alpha = \frac{1}{2}\sum_{n=1}^{N} w_n \mu_n^\alpha \left(\psi_{r,n}^+ + (-1)^\alpha \psi_{r,n}^-\right) \qquad (3.11)$$

where $\alpha = 0$ is for the total intensity, $\alpha = 1$ for the flux and $\alpha = 2$ for the second moment, etc.

## 4. The discrete ordinates transfer equations

The two-streams equations in the *n*th direction on $S_{2N}$ and for $r \in [\varepsilon, R]$ may be written as:

$$\frac{\mu_n}{r^2}\partial_r\left(r^2\psi_n^-\right) + \frac{1}{r}\frac{\delta\left[\eta^2\psi^-\right]_n}{w_n} - \kappa\psi_n^- = -\kappa B \qquad (4.1)$$

$$\frac{\mu_n}{r^2}\partial_r\left(r^2\psi_n^+\right) + \frac{1}{r}\frac{\delta\left[\eta^2\psi^+\right]_n}{w_n} + \kappa\psi_n^+ = \kappa B \qquad (4.2)$$

Define the parameters:

$$\beta_n^n = \frac{\eta_n^2}{w_n\mu_n}; \quad \beta_n^{n-1} = \frac{\eta_{n-1}^2}{w_n\mu_n}; \quad \lambda_n = \frac{\kappa}{\mu_n} \qquad (4.3)$$

These parameters have the following properties:

$$\beta_n^k > 0, \quad \forall n, k \neq N ; \qquad \beta_n^{n-1} > \beta_n^n, \quad \forall n > 1$$
$$\beta_N^N = 0, \quad \forall N ; \quad N \to \infty \Rightarrow \beta_1^1 \to \infty, \quad \beta_N^{N-1} \to 0$$
$$\beta_1^0 \text{ not defined, fill its place by a zero}$$
$$\text{smaller } n \Rightarrow \text{ larger } \beta$$

Make use of identities (4.3) and rearrange (4.1) and (4.2) to read:

$$\partial_r \psi_n^- + \left[\frac{2+\beta_n^n}{r} - \lambda_n\right]\psi_n^- - \frac{1}{r}\beta_n^{n-1}\psi_{n-1}^- = -\lambda_n B \tag{4.4}$$

$$\partial_r \psi_n^+ + \left[\frac{2+\beta_n^n}{r} + \lambda_n\right]\psi_n^+ - \frac{1}{r}\beta_n^{n-1}\psi_{n-1}^+ = \lambda_n B \tag{4.5}$$

*4.1. Matrix representation of the transfer equations*

Define the matrix elements:

$$\mathbf{B}_n^n = (2+\beta_n^n); \quad \mathbf{B}_n^{n-1} = -\beta_n^{n-1}; \quad \Lambda_n^n = \lambda_n \tag{4.6}$$

There should be no confusion between the beta matrix element $\mathbf{B}_n^k$ and the blackbody constant *B*. The latter is italic type font and not indexed. Make use of (4.6) and rewrite (4.5) in matrix form:

$$\partial_r \mathbf{\Psi}^+ = \left(-\frac{1}{r}\mathbf{B} - \Lambda\right)\mathbf{\Psi}^+ + B\boldsymbol{\lambda} = \mathbf{H}\mathbf{\Psi}^+ + B\boldsymbol{\lambda} \tag{4.7}$$

This is the $S_{2N}$ matrix equation on $[\varepsilon, R]$ for the centrifugal intensity. The mathematical boundary condition at the lower end of the domain of definition is $\mathbf{\Psi}_\varepsilon^+$. In general, this boundary condition is not known *a priori*, it will be resolved from the solution of the centripetal equation.

Make use of identities (4.6) and rewrite (4.4) in matrix form:

$$\partial_r \mathbf{\Psi}^- = \left(-\frac{1}{r}\mathbf{B} + \Lambda\right)\mathbf{\Psi}^- - B\boldsymbol{\lambda} = \tilde{\mathbf{H}}\mathbf{\Psi}^- - B\boldsymbol{\lambda} \tag{4.8}$$

This is the $S_{2N}$ matrix equation on $[\varepsilon, R]$ for the centripetal intensity. Its boundary condition is $\mathbf{\Psi}_\varepsilon^-$ which is to be resolved in terms of the natural boundary condition $\mathbf{\Psi}_R^-$ by appropriate inversion of the solution.

The two equations (4.7) and (4.8) merged together yield the matrix equation of the radiative transfer equation defined on $[0, R]$:

$$\partial_r \begin{bmatrix} \mathbf{\Psi}^- \\ \mathbf{\Psi}^+ \end{bmatrix} = \begin{bmatrix} \tilde{\mathbf{H}} & 0 \\ 0 & \mathbf{H} \end{bmatrix} \begin{bmatrix} \mathbf{\Psi}^- \\ \mathbf{\Psi}^+ \end{bmatrix} + B \begin{bmatrix} -\boldsymbol{\lambda} \\ +\boldsymbol{\lambda} \end{bmatrix} \tag{4.9}$$

or, in compact form:

$$\partial_r \mathbf{\Psi} = \mathbf{G}\mathbf{\Psi} + B\boldsymbol{\lambda} \tag{4.10}$$

and the boundary condition is the vector:

$$\Psi_b = \begin{bmatrix} \Psi_\varepsilon^- & \Psi_\varepsilon^+ \end{bmatrix}^T \tag{4.11}$$

The formal solution for (4.10) may be written in the form:

$$\Psi_r = \mathcal{G}(r)\Psi_b + B\int_\varepsilon^r \mathcal{G}(r-x)\lambda\, dx \tag{4.12}$$

where $\mathcal{G}(r)$ is the fundamental matrix of (4.10). It follows that the solution for the transfer equation in the proposed configuration reduces to determining the fundamental matrix $\mathcal{G}$. However, it is less cumbersome and more instructive to solve the centripetal and the centrifugal equations individually in terms of their mathematical boundary conditions then resolve the solutions in terms of the natural boundary condition. This approach is particularly useful in the present case in a medium without scattering. In what follows, the complete solution for the centrifugal equation is carried out and is used to validate the set of discrete ordinates. The complete solution of the two equations is discussed in another paper [11].

## 5. Solutions of the centrifugal equation

### 5.1. Formal representation of the solution

Let $\Gamma$ be the integral of $\mathbf{H}$ on $[\varepsilon, R]$. Then the elements of $\Gamma$ are:

$$\Gamma_n^n = (2+\beta_n^n)\ln\zeta - \lambda_n(r-\varepsilon) \qquad \zeta = \varepsilon/r \tag{5.1}$$

$$\Gamma_n^{n-1} = -\beta_n^{n-1}\ln\zeta$$

It can be shown that the eigenvalues of $\Gamma$ matrix are its diagonal elements and the corresponding eigenvectors are obtained from the recursion relation:

$$\begin{aligned} V_n^n &= 1 & \forall n \\ V_n^k &= 0 & \text{for } k > n \\ V_n^k &= \frac{\Gamma_n^{n-1}}{\Gamma_k^k - \Gamma_n^n} V_{n-1}^k & \text{for } k < n = 2, 3, ..., N \end{aligned} \tag{5.2}$$

It follows that the eigentable $\mathbf{V}_{\varepsilon:r}$ is a lower triangular matrix all its elements are dimensionless. Hence, the solution for the centripetal equation may be cast in the form:

$$\Psi_r^+ = \zeta^2 \mathbf{S}_{\varepsilon:r}\Psi_\varepsilon^+ + \mathbf{Q}_{r:\varepsilon}^+ \lambda \tag{5.3}$$

$$\mathbf{S}_{\varepsilon:r} = \mathbf{V}_{\varepsilon:r}\left[\zeta^{\beta_n^n} e^{-\lambda_n(r-\varepsilon)}\right]\mathbf{V}_{\varepsilon:r}^{-1} \tag{5.4}$$

$$\mathbf{Q}_{r:\varepsilon}^+ = B\int_\varepsilon^r (x/r)^2 \mathbf{S}_{x:r}\, dx \tag{5.4}$$

The disposition of the double indices mirrors the disposition of the same parameters of the ratio $\zeta = \varepsilon/r$ in the spherical operator $\mathbf{S}$ and the limits of integration of the integral of the emission term $\mathbf{Q}$. For the problem at hand, we assume that the radiance $\Psi_\varepsilon^+$ is prescribed. Using the ART properties [2], then the solution (5.3) reads:

$$\Psi_r^+ = \mathbf{T}_r^\varepsilon \Psi_\varepsilon^- + \mathbf{R}_r^\varepsilon \qquad (5.5)$$

$$\mathbf{T}_r^\varepsilon = \zeta^2 \mathbf{S}_{\varepsilon:r}, \qquad \mathbf{R}_r^\varepsilon = \mathbf{Q}_{r:\varepsilon} \lambda$$

The expression (5.5) is the complete solution for the centrifugal equation (2.3) in a spherical shell of thickness $r - \varepsilon : r \in [\varepsilon, R]$ and $R$ can be arbitrarily large. The ART properties $\mathbf{T}_r^\varepsilon$ and $\mathbf{R}_r^\varepsilon$ are the centrifugal transmittance and the centrifugal radiance of the shell, respectively; the transport is from $\varepsilon$ to $r$.

*5.2. The point source problem*

A particular case of the centrifugal intensity is that of a sphere of radius $\varepsilon$ with isotropic and uniform radiance $\psi_\varepsilon$ embedded in an infinite cold absorber, or simply immersed in vacuum. The specific intensity for this problem is given by (5.3) without the emission term:

$$\Psi_r^+ = \zeta^2 \mathbf{S}_{\varepsilon:r} \Psi_\varepsilon^+ \qquad r \geq \varepsilon \qquad (5.6)$$

More often, the sphere is defined as a source of strength $q = 4\pi\varepsilon^2 \psi_\varepsilon$. Then (5.6) reduces to:

$$\Psi_r^+ = \frac{q}{4\pi} \frac{1}{r^2} \mathbf{S}_{\varepsilon:r} \mathbf{1} \qquad r \geq \varepsilon \qquad (5.7)$$

where **1** is an isotropic column vector. Consider the asymptotic case for $r \gg \varepsilon$. For large $N$, all the $\beta$'s are positive except $\beta_N^N = 0$. In this asymptotic limit, $\mathbf{S}_{\varepsilon:r} \to e^{-\kappa r}$, since all the exponentials in $\kappa/\mu$ vanish, or do not contribute to the quadrature for the total intensity. Then the expression (5.7) reduces to:

$$\psi_r = \frac{q}{4\pi} \frac{e^{-\kappa r}}{r^2} \qquad r \gg \varepsilon \qquad (5.8)$$

This expression for the point source problem is reported as a particular case of the flux limited diffusion theory without restriction to the lower bound for $r$ [12]. In the proximity of the surface of the pellet, however small or large $\varepsilon$ may be, either equation (5.6) or (5.7) should be used, whichever applies.

## 6. The stratified medium problem

The continuous-in-$r$ solution of the previous sections is complete in a homogeneous medium. If the medium is not homogeneous but stratified with homogeneous shells, a continuous-in-$r$ solution can be sought for each shell individually. The ensemble of these solutions becomes the solution in the entire sphere if the boundary values for each shell are known.

Consider a sphere made of $M$ homogeneous shells and a pellet. We label the shell contiguous to the pellet with index $m = 1$, and the last shell with $m = M$ as illustrated in figure 6.1. Each shell is characterized by its opacity $\kappa_m$ and a blackbody constant $B_m$. In this configuration, the $m$th stratum is bounded by $r_{m-1}$ and $r_m$. The subscript $r_m$ for the radial position of the $m$th inter-face will be designated by the index $m$ such as $\Psi_m \equiv \Psi(r_m)$. Hence, the *end-points* solution transcribed from (5.5) becomes:

$$\Psi_m^+ = \mathbf{T}_m^{m-1} \Psi_{m-1}^+ + \mathbf{R}_m^{m-1} \qquad m = 1, 2, \ldots, M \tag{6.1}$$

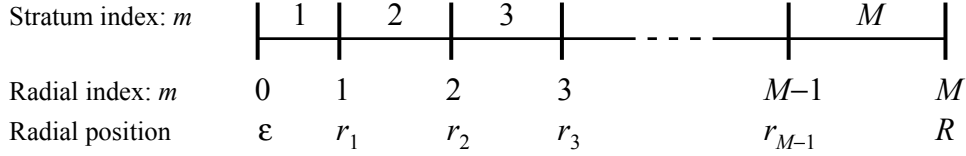

**Figure 6.1.** Labeling of the strata and their boundaries.

This is indeed a numerical algorithm free of spatial truncation errors. It is valid for homogeneous and heterogeneous media. The heterogeneous medium is assumed stratified into $M$ homogeneous shells. In some realistic transfer problems, $r^{-n}$ opacities are more appropriate. In that case, the factor of $\lambda_n$ of (5.1) becomes the integral of $r^{-n}$. We did not implement such opacities in our derivations. Homogeneous strata are sufficient to validate the proposed structure of the set $S_{2N}$. Application of $S_{2N}$ to problems with $r^{-n}$ opacities is discussed in another paper [11].

The specific intensities calculated with the end-points algorithm can be used as boundary values for the continuous-in-$r$ solutions of §5 now defined on the domain $r \in [r_{m-1}, r_m]$. Then (5.3) restricted to the $m$th shell transcribes into:

$$\Psi_r^+ = \zeta_{m-1}^2 \mathbf{S}_r^{m-1} \Psi_{m-1}^+ + \mathbf{Q}_{m-1}^r \lambda_m \tag{6.2}$$

$$\zeta_{m-1} = r_{m-1}/r$$

## 7. Validation of the set of discrete ordinates

The solution we derived in the previous sections is an approximation on the angular domain. The manmade approximation is in the divergence term of the transfer equation, trailing behind it the consequential approximate specific intensity and its moments. The material and the emission components of the solution are not affected by this approximation. Therefore, the validity of the solution with the proposed approximation should be demonstrated for the divergence term alone without emission and interaction with the medium. This condition is met in vacuum. In what follows, we examine the solution in vacuum as well as the import of the approximation to the transfer process in representative absorbing homogeneous media. Vacuum and cold absorbing media are all is needed for the validation of the set $S_{2N}$.

Consider a spherical surface of radius $\varepsilon$ characterized by its isotropic radiance $\psi_\varepsilon^+$ and embedded in a cold medium that can be material of opacity $\kappa$, or eventually vacuum. The specific intensity at point $r \geq \varepsilon$ in a direction defined by $\mu$ is given as:

$$\psi_r^+ = \psi_\varepsilon^+ e^{-\kappa s(\mu)} \tag{7.1}$$

where $s(\mu) = r[\mu - (\zeta^2 - \eta^2)^{1/2}]$ and $\mu \geq \mu_0 = (1 - \zeta^2)^{1/2}$. Integration of (7.1) yields the α-moment at $r$:

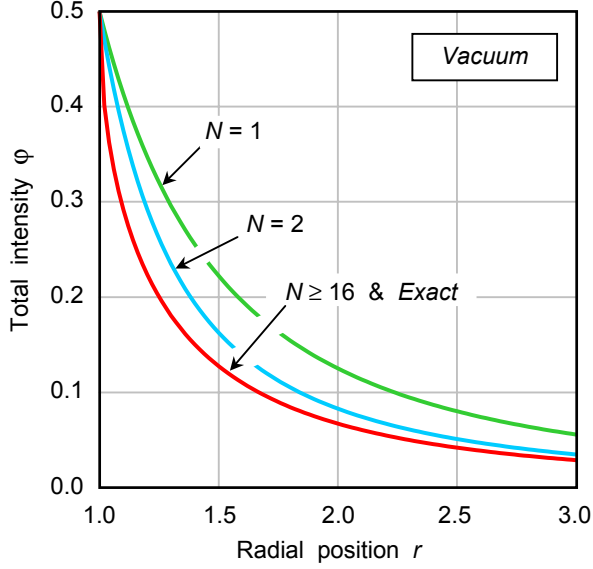

**Figure 7.1**. Graphs of total intensities generated in vacuum with normalized radiance of a sphere of radius $\varepsilon = 1.0$.

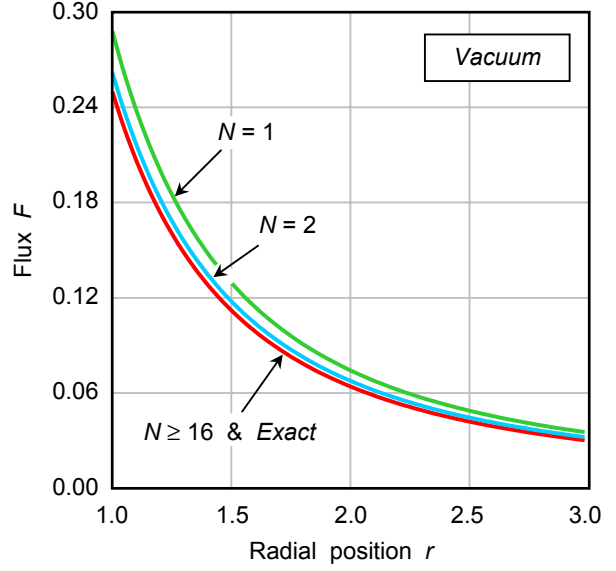

**Figure 7.2**. Graphs of fluxes generated in vacuum with normalized radiance of a sphere of radius $\varepsilon = 1.0$.

in an absorber:
$$f_\alpha = \frac{\psi_\varepsilon^+}{2} \int_{\mu_0}^1 \mu^\alpha e^{-\kappa s(\mu)} d\mu \tag{7.2}$$

and in vacuum:
$$f_\alpha = \frac{\psi_\varepsilon^+}{2(\alpha+1)} \left(1 - \mu_0^{\alpha+1}\right) \tag{7.3}$$

Total intensities calculated using the discrete ordinates solution (6.1) are compared graphically with the total intensities calculated with (7.2) and (7.3). The two sets of data are graphically congruent for $N$ as little as 16 in vacuum and $N = 10$ in cold absorber. The graphs for the total intensity, the flux and the second moment in vacuum are shown in figures 7.1, 7.2 and 7.3, respectively. The label *exact* in these figures refers to the quantities expressed by (7.3). The graphs for the total intensity, the flux and the second moment in an absorber are shown in figures 7.4, 7.5 and 7.6, respectively. The label *exact* in these figures refers to the quantities expressed by (7.2). The integral of (7.2) is calculated using

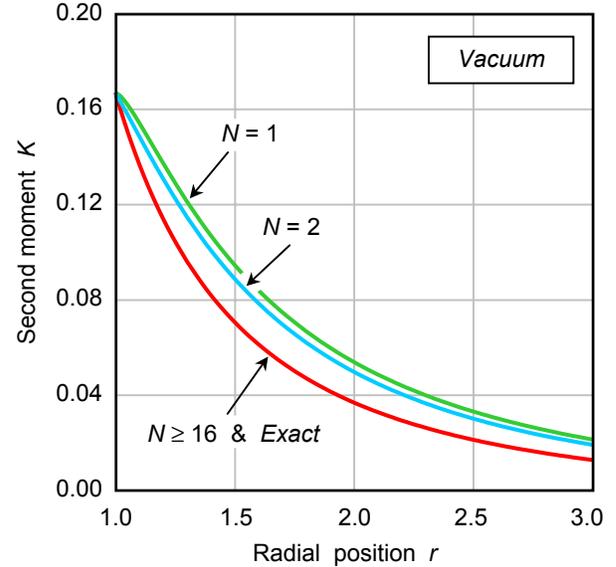

**Figure 7.3**. Graphs of second moments of the specific intensities generated in vacuum with normalized radiance of a sphere of radius $\varepsilon = 1.0$.

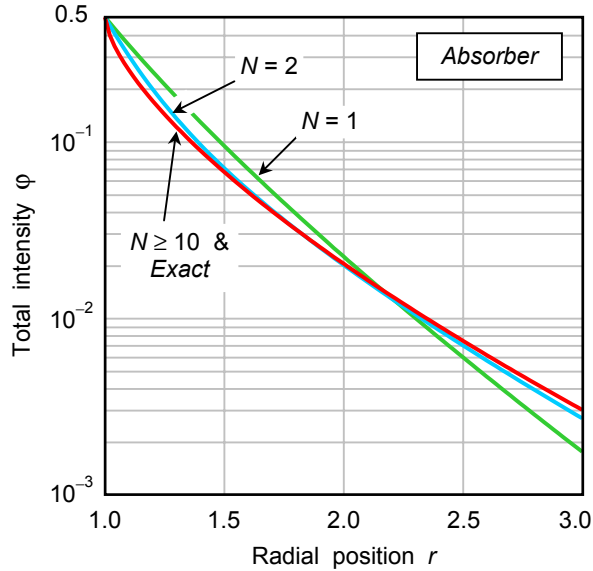

**Figure 7.4**. Graphs of total intensities generated in a cold homogeneous medium of opacity $\kappa = 1$ with normalized radiance of a sphere of radius $\varepsilon = 1.0$.

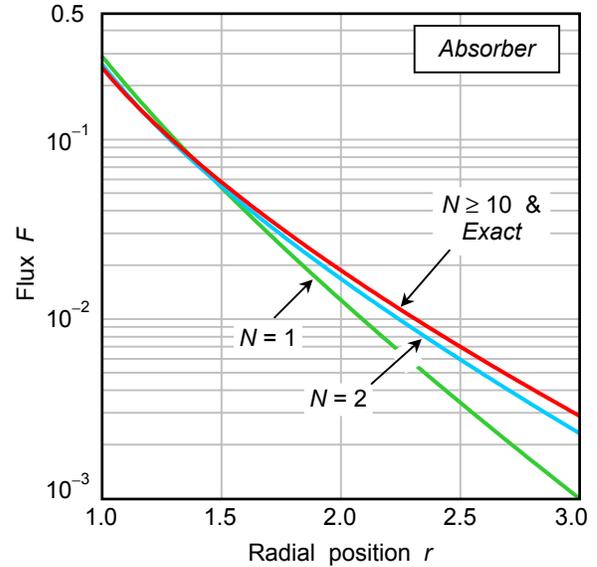

**Figure 7.5**. Graphs of fluxes generated in a cold homogeneous medium of opacity $\kappa = 1$ with normalized radiance of a sphere of radius $\varepsilon = 1.0$.

Rhomberg algorithm. The extent of the graphs to $\zeta = 1/3$ is deliberate for the interest of showing details of the discrepancies particularly near the surface of the sphere. Other values of $\zeta$ were tested. The graphical congruence was achieved with values of $N = 10$.

The graphs of the total intensities in these figures are smooth monotonically decreasing quite as expected. Their degenerate specific intensities are not so. Figure 7.7 shows the graphs of specific intensities generated in vacuum in different directions defined by $\mu_n$ extracted from the set of $N = 40$ discrete ordinates; the graph of the total intensity is shown in the same figure. The intensity $\psi_{N=40}$, which corresponds to $\mu_N$, is nominally emitted normally to the surface of the radiant pellet, yet this intensity peaks to a little more than twice its surface value before it starts to decrease monotonically. The peak gets smaller and closer to the surface of the pellet with decreasing $n$. This behavior is attributable to the coupling of the intensities by the angular derivative: specific intensities in slant directions contribute to the intensity in the radial direction.

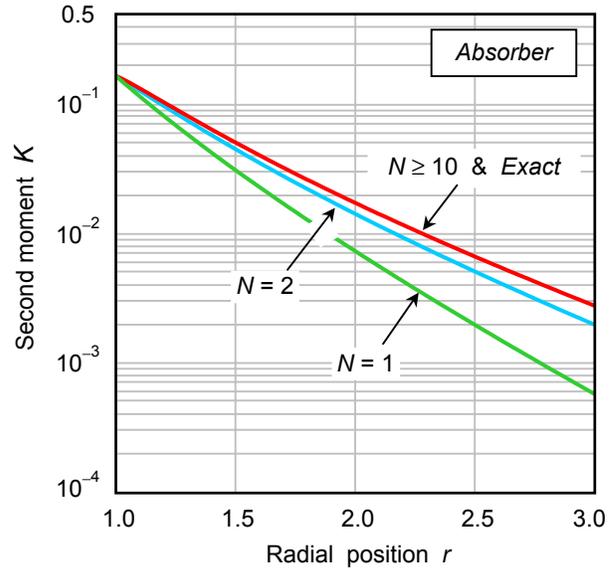

**Figure 7.6**. Graphs of the second moments generated in a cold homogeneous medium of opacity $\kappa = 1$ with normalized radiance of a sphere of radius $\varepsilon = 1.0$.

The pattern of these graphs is generic. It was observed with all radiating pellets and for all orders of discrete ordinates we experimented with; the magnitudes of the intensities are smaller in absorbing media by reason of attenuation. The graphs of figure 7.7 are generated with the discrete ordinates solution (6.1). Their corresponding graphs using (7.1) are a horizontal line for all directions. This observation leads us to conclude that the pattern of the specific intensity of figure 7.7 is peculiar to GL quadrature. Considering that different quadratures have different weights and abscissas, the pattern of figure 7.7 may not necessarily be representative of specific intensities computed with different quadratures.

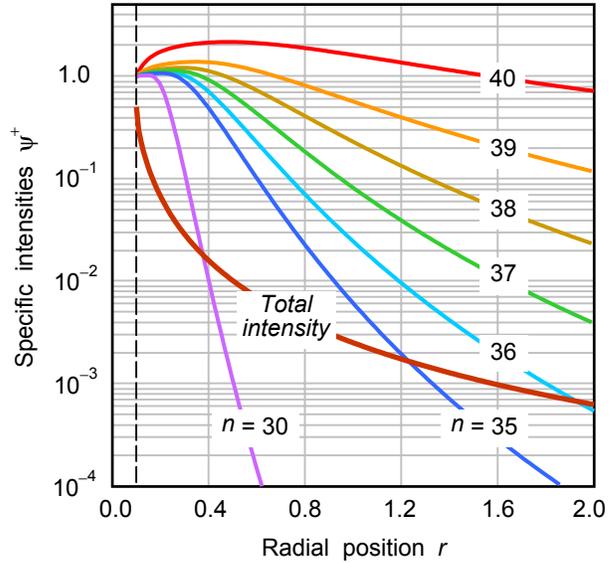

**Figure 7.7**. Graphs of the total intensity and some of its angular degenerates specific intensities in $N = 40$ discrete ordinates in vacuum. They are generated from a pellet of radius $\varepsilon = 0.1$.

We realize that the total intensities computed in discrete ordinates and the one computed using the analytic expressions (7.1) must be, and they are in the present case identically the same. This observation leads us to conclude that the total intensity and eventually the higher moments are invariant under quadrature if the set of discrete ordinates is properly structured. We cannot infer this invariance property to specific intensities.

The reading of figure 7.7 suggests a way, however simplistic it may be, to visualize graphically the variation of the radiation field in the interior of a spherical cavity. Let $R = 2$ be the surface of a cavity with a concentric reflective pellet and let $\psi_R^-$ be the intensity of the external radiation field. If we preserve the integrity of the principle of reciprocity of path and we choose $\psi_R^-$ such that the intensity at the surface of the pellet is isotropic, as illustrated in figure 7.7, then the external radiation field becomes physically unacceptable. On the other hand, if $\psi_R^-$ is isotropic, then the radiation field becomes nonphysical most everywhere in the cavity.

This naïve attempt to construct the solution in the spherical cavity lacks the effect of the superposition of the contributions from the angular derivatives of the two-streams intensities. The culprit is not the discrete ordinates. It is the unfiltered implementation of the angular derivative in the two-streams radiation field. The proper implementation of the angular derivative in the interior of spheres is not transparent in the analysis of this paper. It is discussed elsewhere [11].

Tabulated data were also generated in vacuum for macrometric and micrometric pellets and for different orders of discrete ordinates $N$. Samples of these data are shown in Table 7.1 for a spherical region bounded by $\varepsilon = 10^{-7}$ and $R = 1.0$, and in Table 7.2 for a spherical region bounded by $\varepsilon = 1.0$ and $R = 10^7$. The data in these tables are generated with 200 spatial intervals. Larger and smaller number of spatial intervals reproduced the same numbers. This is expected since the solutions are exact in $r$.

**Table 7.1.** Computed differences for the total intensities $\Delta\varphi$, fluxes $\Delta F$ and second moments $\Delta K$ obtained from the exact values and the discrete ordinates values; the conservation of energy $E$ is computed in discrete ordinates. Various orders of discrete ordinates $N$ are considered at different radial positions. The source is a macropellet of radius $\varepsilon = 1.0$ its radiance is normalized to 1.

| | Radial position $R$ = 1 to $10^7$ in vacuum media | | | | | | |
|---|---|---|---|---|---|---|---|
| $N$ | 1 | 2 | 4 | 10 | $10^2$ | $10^4$ | $10^7$ |
| | Total intensity $\Delta\varphi$ | | | | | | |
| 10 | $-5.5\times10^{-17}$ | $1.3\times10^{-3}$ | $2.5\times10^{-4}$ | $2.9\times10^{-5}$ | $2.2\times10^{-7}$ | $2.2\times10^{-11}$ | $2.4\times10^{-17}$ |
| 20 | $-1.1\times10^{-16}$ | $5.6\times10^{-4}$ | $9.8\times10^{-5}$ | $1.0\times10^{-5}$ | $5.9\times10^{-8}$ | $5.6\times10^{-12}$ | $7.6\times10^{-18}$ |
| 30 | $-1.6\times10^{-16}$ | $3.9\times10^{-4}$ | $5.9\times10^{-5}$ | $5.7\times10^{-6}$ | $2.7\times10^{-8}$ | $2.5\times10^{-12}$ | $4.5\times10^{-18}$ |
| 40 | $1.1\times10^{-16}$ | $2.5\times10^{-4}$ | $4.1\times10^{-5}$ | $3.9\times10^{-6}$ | $1.6\times10^{-8}$ | $1.4\times10^{-12}$ | $3.4\times10^{-18}$ |
| 50 | $-1.1\times10^{-16}$ | $2.1\times10^{-4}$ | $3.9\times10^{-5}$ | $3.3\times10^{-6}$ | $1.4\times10^{-8}$ | $1.2\times10^{-12}$ | $3.2\times10^{-18}$ |
| | Flux $\Delta F$ | | | | | | |
| 10 | $4.9\times10^{-4}$ | $1.2\times10^{-4}$ | $3.0\times10^{-5}$ | $4.9\times10^{-6}$ | $4.9\times10^{-8}$ | $4.9\times10^{-12}$ | $6.9\times10^{-18}$ |
| 20 | $1.2\times10^{-4}$ | $3.1\times10^{-5}$ | $7.8\times10^{-6}$ | $1.2\times10^{-6}$ | $1.2\times10^{-8}$ | $1.2\times10^{-12}$ | $3.2\times10^{-18}$ |
| 30 | $5.6\times10^{-5}$ | $1.4\times10^{-5}$ | $3.5\times10^{-5}$ | $5.6\times10^{-7}$ | $5.6\times10^{-9}$ | $5.6\times10^{-13}$ | $2.5\times10^{-18}$ |
| 40 | $3.1\times10^{-5}$ | $7.8\times10^{-6}$ | $1.9\times10^{-6}$ | $3.1\times10^{-7}$ | $3.1\times10^{-9}$ | $3.1\times10^{-13}$ | $2.3\times10^{-18}$ |
| 50 | $2.0\times10^{-5}$ | $2.3\times10^{-5}$ | $4.0\times10^{-6}$ | $5.4\times10^{-7}$ | $5.4\times10^{-9}$ | $5.4\times10^{-13}$ | $2.5\times10^{-18}$ |
| | Second moment $\Delta K$ | | | | | | |
| 10 | $2.7\times10^{-17}$ | $-6.4\times10^{-4}$ | $-1.6\times10^{-4}$ | $-1.8\times10^{-5}$ | $-1.2\times10^{-7}$ | $-1.2\times10^{-11}$ | $-1.0\times10^{-17}$ |
| 20 | $-5.5\times10^{-17}$ | $-2.8\times10^{-4}$ | $-7.2\times10^{-5}$ | $-7.3\times10^{-6}$ | $-3.4\times10^{-8}$ | $-3.1\times10^{-12}$ | $-1.1\times10^{-18}$ |
| 30 | $-1.9\times10^{-16}$ | $-1.8\times10^{-4}$ | $-4.5\times10^{-5}$ | $-4.5\times10^{-6}$ | $-1.6\times10^{-8}$ | $-1.4\times10^{-12}$ | $5.8\times10^{-19}$ |
| 40 | $8.3\times10^{-17}$ | $-1.3\times10^{-4}$ | $-3.3\times10^{-5}$ | $-3.2\times10^{-6}$ | $-9.8\times10^{-9}$ | $-7.9\times10^{-13}$ | $1.2\times10^{-18}$ |
| 50 | $-8.3\times10^{-17}$ | $-8.5\times10^{-5}$ | $-2.3\times10^{-5}$ | $-2.6\times10^{-7}$ | $-3.3\times10^{-9}$ | $-1.7\times10^{-13}$ | $1.8\times10^{-18}$ |
| | Energy balance $E$ | | | | | | |
| 10 | 0.0 | $-1.7\times10^{-15}$ | $-1.1\times10^{-16}$ | $2.5\times10^{-16}$ | $4.9\times10^{-16}$ | $1.6\times10^{-16}$ | $4.2\times10^{-16}$ |
| 20 | 0.0 | $-7.9\times10^{-13}$ | $-3.9\times10^{-13}$ | $-1.5\times10^{-13}$ | $-1.8\times10^{-13}$ | $-1.8\times10^{-13}$ | $-1.8\times10^{-13}$ |
| 30 | 0.0 | $-3.8\times10^{-10}$ | $-2.4\times10^{-11}$ | $1.0\times10^{-10}$ | $1.5\times10^{-10}$ | $1.5\times10^{-10}$ | $1.5\times10^{-10}$ |
| 40 | 0.0 | $2.0\times10^{-7}$ | $1.0\times10^{-7}$ | $4.8\times10^{-8}$ | $3.3\times10^{-8}$ | $3.3\times10^{-8}$ | $3.3\times10^{-8}$ |
| 50 | 0.0 | $-7.3\times10^{-5}$ | $-4.3\times10^{-5}$ | $-3.4\times10^{-5}$ | $-3.4\times10^{-5}$ | $-3.4\times10^{-5}$ | $-3.4\times10^{-5}$ |

The quantity $\Delta\varphi = \varphi_{discr} - \varphi_{exact}$ represents the difference between the total intensity computed in discrete ordinates with (6.1) and the total intensity using the exact values (7.3). Similarly, $\Delta F$ is the difference between the discrete ordinates and the exact fluxes, and $\Delta K$ the difference between the discrete ordinates and the exact second moments. All quantities, $\varphi$, $F$ and $K$ of the tables are computed from a radiance $\psi_\varepsilon^+$ normalized to 1.0 using 16 significant digits arith-metics (16-SDA). Considering that in vacuum the **S**-operator is function of $\zeta$ only, one should expect the columns that correspond to the same $\zeta$ to be identically the same in both of the tables. We maintained $N \leq 50$ in vacuum to avoid non-negligible round-off errors.

**Table 7.2.** Computed differences for the total intensities $\Delta\varphi$, fluxes $\Delta F$ and second moments $\Delta K$ obtained from the exact values and the discrete ordinates values; the conservation of energy $E$ is computed in discrete ordinates. Various orders of discrete ordinates $N$ are considered at different radial positions. The source is a micropellet of radius $\varepsilon = 10^{-7}$ its radiance is normalized to 1.

| | Radial position $R = 10^{-7}$ to 1.0 in vacuum media | | | | | | |
|---|---|---|---|---|---|---|---|
| $N$ | $10^{-7}$ | $10^{-6}$ | $10^{-4}$ | $10^{-2}$ | $10^{-1}$ | 0.5 | 1.0 |
| | Total intensity $\Delta\varphi$ | | | | | | |
| 10 | $-5.5\times10^{-17}$ | $2.9\times10^{-5}$ | $2.2\times10^{-9}$ | $2.2\times10^{-13}$ | $2.2\times10^{-15}$ | $9.6\times10^{-17}$ | $2.4\times10^{-17}$ |
| 20 | $-1.1\times10^{-16}$ | $1.0\times10^{-5}$ | $5.6\times10^{-10}$ | $5.6\times10^{-14}$ | $5.4\times10^{-16}$ | $3.0\times10^{-17}$ | $7.7\times10^{-18}$ |
| 30 | $-1.6\times10^{-16}$ | $5.7\times10^{-6}$ | $2.5\times10^{-10}$ | $2.5\times10^{-14}$ | $2.3\times10^{-16}$ | $1.8\times10^{-17}$ | $5.5\times10^{-18}$ |
| 40 | $1.1\times10^{-16}$ | $3.9\times10^{-6}$ | $1.4\times10^{-10}$ | $1.4\times10^{-14}$ | $1.2\times10^{-16}$ | $1.3\times10^{-17}$ | $3.4\times10^{-18}$ |
| 50 | $-1.1\times10^{-16}$ | $3.3\times10^{-6}$ | $1.2\times10^{-10}$ | $1.2\times10^{-14}$ | $1.0\times10^{-16}$ | $1.3\times10^{-17}$ | $3.2\times10^{-18}$ |
| | Flux $\Delta F$ | | | | | | |
| 10 | $4.9\times10^{-4}$ | $4.9\times10^{-6}$ | $4.9\times10^{-10}$ | $4.9\times10^{-14}$ | $4.6\times10^{-16}$ | $2.7\times10^{-17}$ | $6.9\times10^{-18}$ |
| 20 | $1.2\times10^{-4}$ | $1.2\times10^{-6}$ | $1.2\times10^{-10}$ | $1.2\times10^{-14}$ | $1.0\times10^{-16}$ | $1.3\times10^{-17}$ | $3.2\times10^{-18}$ |
| 30 | $5.6\times10^{-5}$ | $5.6\times10^{-7}$ | $5.6\times10^{-11}$ | $5.6\times10^{-15}$ | $3.4\times10^{-17}$ | $1.0\times10^{-17}$ | $2.5\times10^{-18}$ |
| 40 | $3.1\times10^{-5}$ | $3.1\times10^{-7}$ | $3.1\times10^{-11}$ | $3.1\times10^{-15}$ | $9.4\times10^{-18}$ | $9.6\times10^{-18}$ | $2.3\times10^{-18}$ |
| 50 | $2.0\times10^{-5}$ | $5.5\times10^{-7}$ | $5.4\times10^{-11}$ | $5.4\times10^{-15}$ | $3.2\times10^{-17}$ | $1.0\times10^{-17}$ | $2.5\times10^{-18}$ |
| | Second moment $\Delta K$ | | | | | | |
| 10 | $2.7\times10^{-17}$ | $-1.8\times10^{-5}$ | $-1.2\times10^{-9}$ | $-1.2\times10^{-13}$ | $-1.2\times10^{-15}$ | $-5.2\times10^{-17}$ | $-1.0\times10^{-17}$ |
| 20 | $-5.5\times10^{-17}$ | $-7.3\times10^{-6}$ | $-3.1\times10^{-10}$ | $-3.1\times10^{-14}$ | $-3.3\times10^{-16}$ | $-4.6\times10^{-18}$ | $-1.1\times10^{-18}$ |
| 30 | $-1.9\times10^{-16}$ | $-4.5\times10^{-6}$ | $-1.4\times10^{-10}$ | $-1.4\times10^{-14}$ | $-1.6\times10^{-16}$ | $2.3\times10^{-18}$ | $5.8\times10^{-19}$ |
| 40 | $8.3\times10^{-17}$ | $-3.2\times10^{-6}$ | $-8.0\times10^{-11}$ | $-7.9\times10^{-15}$ | $-1.0\times10^{-16}$ | $4.8\times10^{-18}$ | $1.2\times10^{-18}$ |
| 50 | $-8.3\times10^{-17}$ | $-2.1\times10^{-6}$ | $-1.7\times10^{-11}$ | $-1.7\times10^{-15}$ | $-3.9\times10^{-17}$ | $7.3\times10^{-18}$ | $1.8\times10^{-18}$ |
| | Energy balance $E$ | | | | | | |
| 10 | 0.0 | $1.2\times10^{-30}$ | $-3.4\times10^{-30}$ | $9.7\times10^{-30}$ | $6.1\times10^{-31}$ | $1.1\times10^{-29}$ | $5.2\times10^{-30}$ |
| 20 | 0.0 | $-1.6\times10^{-27}$ | $-1.8\times10^{-27}$ | $-1.8\times10^{-27}$ | $-1.8\times10^{-27}$ | $-1.8\times10^{-27}$ | $-1.8\times10^{-27}$ |
| 30 | 0.0 | $1.0\times10^{-24}$ | $1.5\times10^{-24}$ | $1.5\times10^{-24}$ | $1.5\times10^{-24}$ | $1.5\times10^{-24}$ | $1.5\times10^{-24}$ |
| 40 | 0.0 | $4.7\times10^{-22}$ | $3.3\times10^{-22}$ | $3.3\times10^{-22}$ | $3.3\times10^{-22}$ | $3.3\times10^{-22}$ | $3.3\times10^{-22}$ |
| 50 | 0.0 | $-3.4\times10^{-19}$ | $-3.4\times10^{-19}$ | $-3.4\times10^{-19}$ | $-3.4\times10^{-19}$ | $-3.4\times10^{-19}$ | $-3.4\times10^{-19}$ |

The quantity $E$ in the tables represents the energy balance. It is obtained by integrating (2.1) over all directions and over the volume of interest: $E = R^2 F_R - \varepsilon^2 F_\varepsilon + \kappa\varphi_{int}$, where $\varepsilon^2 F_\varepsilon$ represents the total energy radiating out of the pellet into the shell, $R^2 F_R$ the total energy radiating out of the shell, and $\kappa\varphi_{int}$ the total energy absorbed in the volume of the shell, which is zero in vacuum space; $\varphi_{int}$ is the integral of the total intensity obtained numerically using Simpson's 3/8 rule,

**Table 7.3**. Tabulated data for the total intensity φ, the flux $F$, and the second moment $K$, at the surface of a radiating pellet of radius ε = 1 and at radial positions $R_3$ and $R_5$, and the energy balance $E$ in the volume of the shell of thickness $R_5 - ε$. The data are computed in $N$ discrete ordinates and using the exact expression in gray atmospheres of opacities κ = 1 and 4.

| | ε = 1.0, $R_3$ = 3.0, $R_5$ = 5.0, κ = 1.0 | | | | | | | | |
|---|---|---|---|---|---|---|---|---|---|
| N | $φ_1$ | $φ_3×10^3$ | $φ_5×10^4$ | $F_1$ | $F_3×10^3$ | $F_5×10^4$ | $K_1$ | $K_3×10^3$ | $K_5×10^4$ | $E×10^7$ |
| 10 | 0.5 | 3.0036 | 1.3639 | 0.25049 | 2.8940 | 1.3389 | 1.66 | 2.7934 | 1.3150 | 6.79 |
| 20 | 0.5 | 3.0346 | 1.3965 | 0.25012 | 2.9433 | 1.3788 | 1.66 | 2.8575 | 1.3617 | 6.96 |
| 30 | 0.5 | 3.0439 | 1.4059 | 0.25006 | 2.9576 | 1.3901 | 1.66 | 2.8757 | 1.3747 | 7.03 |
| 40 | 0.5 | 3.0484 | 1.4102 | 0.25003 | 2.9642 | 1.3953 | 1.66 | 2.8841 | 1.3807 | 6.39 |
| 50 | 0.5 | 3.0513 | 1.4128 | 0.25002 | 2.9683 | 1.3984 | 1.66 | 2.8893 | 1.3843 | 22.6 |
| exact | 0.5 | 3.0607 | 1.4214 | 0.25000 | 2.9819 | 1.4086 | 1.66 | 2.9060 | 1.3960 | - |

| | ε = 1.0, $R_3$ = 3.0, $R_5$ = 5.0, κ = 4.0 | | | | | | | | |
|---|---|---|---|---|---|---|---|---|---|
| N | $φ_1$ | $φ_3×10^6$ | $φ_5×10^{10}$ | $F_1$ | $F_3×10^6$ | $F_5×10^{10}$ | $K_1$ | $K_3×10^6$ | $K_5×10^{10}$ | $E×10^9$ |
| 10 | 0.5 | 4.2216 | 4.3841 | 0.25049 | 4.1175 | 4.3289 | 1.66 | 4.0192 | 4.2753 | −4.89 |
| 20 | 0.5 | 4.3552 | 4.6760 | 0.25012 | 4.2649 | 4.6366 | 1.66 | 4.1787 | 4.5980 | −2.89 |
| 30 | 0.5 | 4.3873 | 4.7461 | 0.25006 | 4.3002 | 4.7105 | 1.66 | 4.2166 | 4.6755 | 9.76 |
| 40 | 0.5 | 4.4004 | 4.7747 | 0.25003 | 4.3146 | 4.7407 | 1.66 | 4.2320 | 4.7073 | 52.3 |
| 50 | 0.5 | 4.4072 | 4.7895 | 0.25002 | 4.3220 | 4.7564 | 1.66 | 4.2399 | 4.7237 | 3939 |
| exact | 0.5 | 4.4264 | 4.8295 | 0.25000 | 4.3426 | 4.7988 | 1.66 | 4.2614 | 4.7683 | - |

claimed to be accurate to $O(h^5 f^{(4)})$. Ideally, $E = 0$.

The information we obtain from the analysis of the data of the tables is perhaps most instructive in the block Δ$F$. The flux in this geometry varies rigorously as $1/r^2$ from the surface of the pellet forward. Therefore, one would expect that Δ$F$ varies as $1/r^2$ within round-off errors. This appears to be truly the case in the two tables. On the other hand, the values of Δ$F$ are different from zero but they get closer to zero in the asymptotic region. This is primarily because at the surface of the pellet, the first column of the tables, we have Δ$F_ε$ = 0.25 − 0.5×Σ$w_i μ_i$. This value cannot be zero except in the limit as $N → ∞$. Indeed, for the particular case of $N = 50$ of the tables, Σ$w_i μ_i$ = 0.5000407 computed in 16-SDA.

Unlike the case of the flux, Δφ at the surface of the pellet is practically zero. This is because of the way the weights of the quadrature are calculated: Σ$w_i$ = 1.0 − 2×10$^{-16}$ in 16-SDA. However, a look at the data of the second column reveals that the total intensity computed in discrete ordinates is noticeably larger than the one computed with the exact expression (7.3). This difference is attributable to the coupling of the intensities by the angular derivative. Finer angular grid reduces the value of Δφ slowly with increasing $N$. The rate of convergence appears to be slow in this case.

The fluxes and total intensities needed to calculate $E$ are all computed in discrete ordinates. The values of $E$ appear to be uniform with increasing $R$ and for the same value of $N$, but the magnitudes of the data of Table 7.1 are closely 14 orders of magnitude larger than the ones of Table 7.2. This is understood from the fact that $F_ε$ is multiplied by $ε^2 = 10^{-14}$ at the surface of the micro-pellet (Tab. 7.2) and by 1.0 at the surface of the macro-pellet (Tab. 7.1), and $F_R$ is multiplied by $R^2 = 1.0$ in the case of micro-pellet and by $10^{14}$ in the case of macro-pellet. It appears, however, that the value of $E$ increases with increasing $N$ in both micro- and macro-pellets.

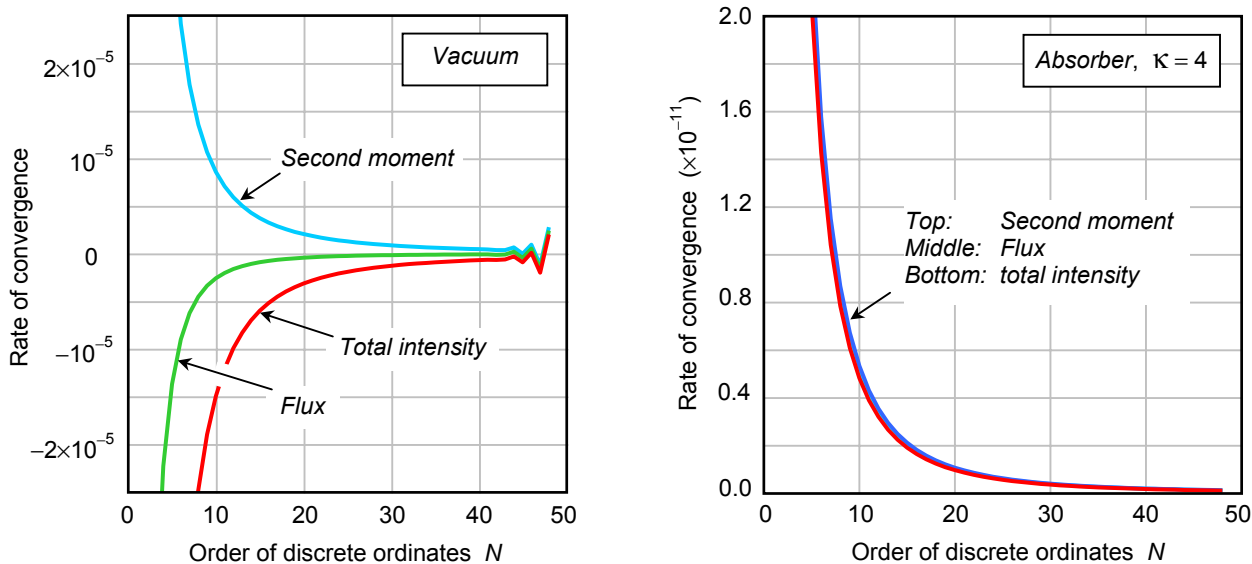

**Figure 7.8**. Graphs of the rate of convergence with respect to the order of discrete ordinates $N$ of the computed total intensity, flux and second moment at $R = 5$ for a radiant pellet of radius $\varepsilon = 1$ and for two different media.

The increase in $E$, noticeable in both of the tables, is directly attributable to round-off errors in matrix operations: larger $N$ generates larger matrices resulting with larger round-off errors. This can be visualized by this sample of data computed in $N = 50$ in the macro-pellet configuration (Tab. 7.1): $\varepsilon^2 F_\varepsilon = 0.25002$ and $R^2 F_R = 0.25005$. The difference 0.00003 results from round-off errors. By contrast, in $N = 10$ computations, these numbers differ in the 13th significant digit: $\varepsilon^2 F_\varepsilon = 0.2504...018$, $R^2 F_R = 0.2504...017$. These figures explain the increase in $E$ as $N$ increases from 10 to 50. Notwithstanding, the computed physical quantities get closer to the exact values with increasing $N$ up to $N = 50$. Past this limit, the round-off error becomes overwhelming in vacuum media.

We collected data in absorptive media 5 and 20 mfp thick These data are shown in Table 7.3. They are computed with $N$ discrete ordinates and with the exact expression (7.2) at different radial positions: at $\varepsilon = 1$, $R_5 = 5$ and at a surface halfway between $\varepsilon$ and $R_5$, which corresponds to $R_3 = 3$. It appears that the values of $\varphi$, $F$, $K$ and $E$ increase with increasing $N$ following a pattern similar to the one observed in the case of vacuum media, suggesting a convergence toward the exact values. The tabulated data in the rows labeled *exact* are obtained with 8,192 evaluations of the integrand of (7.2) using Romberg algorithm. Not attempts were made to interpolate data at this stage. The raw data is intended to reveal the performance of the method in its native form.

The rate of convergence $\Delta f_\alpha/\Delta N$ computed in discrete ordinates for $\Delta N = 1$, is depicted graphically by figure 7.8 as a function of $N$ and at the surface $R = 5$, in vacuum and in a pure absorber. It is apparent that the function $\Delta f$ is smooth and monotonically decreasing but slowly converging to zero in both of the media, vacuum and absorber. However, wriggles in the graphs in vacuum start to develop as $N$ increases past 40. These fluctuations are symptoms of round-off errors. They are small for $N \leq 50$ but they render the data useless for $N > 52$ in vacuum. The absorber medium flattens out these wriggles and the convergence is smooth but very slow all the way through $N = 50$; we have not experimented with $N > 52$. The flux appears to converge the fastest, followed by the second moment then by the total intensity in vacuum. In the absorber medium, the total intensity appears to converge the fastest, followed by the flux then by the second moment. However, the rate of convergence in this medium appears to be closely the same for all the three quantities and for all $N$.

## 8. Summary and conclusions

We structured a new set of discrete ordinates that is valid in one-dimensional spherical geometry radiative transfer problem. The set is built from un-normalized circular functions. This structure allows simple discrete representation of the angular derivative of the conservation form of the transfer equation. This is made with a set of angular parameters obtained directly and without recursion from the abscissas and weights of the quadrature and they are not constrained to asymptotic condition. The new set is closed: it produces $2N$ equations with $2N$ unknowns and requires not a starter intensity. These equations satisfy the conservation relation $E$, rigorously in analytic form, within round off errors in numerical computations.

The features of the proposed set of discrete ordinates are quite in contrast with the traditional discrete ordinates [3]. The latter is structured around a set of angular parameters that are computed recursively from the weights and abscissas of GL quadrature, in the one-dimensional geometry. These angular parameters approximate the angular derivative and they are constrained to an isotropic specific intensity. This structure yields an open set of $2N$ equations with $2N + 1$ unknowns. The additional unknown, labeled the starter intensity, is borrowed from the solution in slab geometry.

We obtained the closed form analytic solution in $r$ for the discrete ordinates centrifugal equation in an infinite cold homogeneous medium and presented the solution in two forms: the continuous-in-$r$ solution and the end-points solution. The latter is an effective numerical algorithm free of spatial truncation errors and is valid in all homogeneous and non-homogeneous extended media. Graphical and tabulated data obtained with the proposed discrete ordinates model appear to agree very well with the data obtained from exact expressions for the total intensity, the flux and the second moment in infinite media.

We restricted our analysis in this paper to the one-dimensional spherical geometry and to vacuum and cold media where the intensity and its moments are exclusively centrifugal. This is in line with the object of this paper: to construct a set of discrete ordinates and to demonstrate the validity of the proposed structure. We demonstrated that effectively. The graphical illustrations and tabulated data demonstrate that the proposed new structure of the set of discrete ordi-nates is sound. It promises high accuracy in radiative transfer computations in the point source configuration where the energy transfer is exclusively centrifugal and the radiation field is highly peaked forward. It is our expectation that the performance of this discrete ordinates model will be significantly better in media with scattering. This is because emission and scattering tend to smoothen out small fluctuations in approximations of the angular derivative, a well-known phenomenon in discrete ordinates computations.

We made no efforts in this paper to collect data with emission from distributed sources, or in media with scattering. Emission and scattering generate a set of two-streams intensities that compete in setting the radiation field in the medium. Similarly, the radiation field in the interior of a cold sphere exposed to a uniformly distributed external radiation field is set up by competing two streams of specific intensities by reason of symmetry.

We illustrated a way in a previous section to visualize the spatial variation of a radiation field set up by two streams of intensities in a spherical cavity; the intensities were computed using the proposed set of discrete ordinates. That graphical experiment led to non-physical intensities everywhere in the cavity. That experiment is tantamount to a graphical inversion of the **S** operator as if it were an algebraic matrix. The spherical operators for either the centripetal or the centrifugal intensity are indeed differential operators on the set $S_{2N}$. Simple algebraic matrix inversion of these operators is most inappropriate. The proper inversion of these operators yields a radiation field that is physically acceptable everywhere in the sphere. This problem is discussed in another paper [11].

We contemplate expanding the proposed structure of the set of discrete ordinates to the three-dimensional spherical geometry in the future.

**Acknowledgment.** This work was funded by the Department of Navy, grant N00173-001-G010.